\documentclass[12pt]{article}
\usepackage{a4wide}
\usepackage{latexsym}
\usepackage{axodraw}
\usepackage{amsmath}
\usepackage{amsfonts}
\usepackage{amscd}
\usepackage{cite}

\usepackage{pslatex}
\usepackage{graphicx}
\usepackage[latin1]{inputenc}
\usepackage[T1]{fontenc}

\newcommand{\bq}{\begin{eqnarray}}
\newcommand{\eq}{\end{eqnarray}}
\newcommand{\eps}{\varepsilon}

\begin{document}

\thispagestyle{empty}

\begin{flushright}
  MZ-TH/12-58 
\end{flushright}

\vspace{1.5cm}

\begin{center}
  {\Large\bf Picard-Fuchs equations for Feynman integrals\\
  }
  \vspace{1cm}
  {\large Stefan M\"uller-Stach ${}^{a}$, Stefan Weinzierl ${}^{b}$ and Raphael Zayadeh ${}^{a}$\\
  \vspace{1cm}
      {\small ${}^{a}$ \em PRISMA Cluster of Excellence, Institut f{\"u}r Mathematik, Universit{\"a}t Mainz,}\\
      {\small \em D - 55099 Mainz, Germany}\\
  \vspace{2mm}
      {\small ${}^{b}$ \em PRISMA Cluster of Excellence, Institut f{\"u}r Physik, Universit{\"a}t Mainz,}\\
      {\small \em D - 55099 Mainz, Germany}\\
  } 
\end{center}

\vspace{2cm}

\begin{abstract}\noindent
  {
We present a systematic method to derive an ordinary differential equation for any Feynman integral, where the differentiation is with
respect to an external variable.
The resulting differential equation is of Fuchsian type.
The method can be used within fixed integer space-time dimensions as well as within dimensional regularisation.
We show that finding the differential equation is equivalent to solving a linear system of equations.
We observe interesting factorisation properties of the $D$-dimensional Picard-Fuchs operator when $D$ is specialised to integer dimensions.
   }
\end{abstract}

\vspace*{\fill}

\newpage

\section{Introduction}

Differential equations are a powerful tool to compute 
Feynman integrals \cite{Kotikov:1990kg,Kotikov:1991pm,Remiddi:1997ny,Gehrmann:1999as,Gehrmann:2000zt,Gehrmann:2001ck,Argeri:2007up}.
Up to now, the standard way to derive a set of differential equations for a set of unknown Feynman integrals is based on
integration-by-parts identities \cite{Tkachov:1981wb,Chetyrkin:1981qh}.
This usually results in a system of coupled differential equations for a set of so-called master integrals.
In addition, there is no guarantee that the resulting system of differential equations is of minimal order.
In fact, a counter-example has been found recently \cite{MullerStach:2011ru}: The two-loop sunrise integral with three unequal masses in two space-time
dimensions satisfies an ordinary second-order differential equation. 
Integration-by-parts identities lead only to a coupled system of four first-order differential equations.

We are interested in finding for a given Feynman integral and a given external variable an ordinary differential equation of minimal order.
In this paper we present an algorithm to solve this problem.
Our method works for arbitrary space-time dimensions and can thus be used with dimensional regularisation.
The method reduces the problem of finding an ordinary differential equation of order $r$ to the task of solving a system of linear equations.
In order to find the differential operator of minimal order, one starts with $r=1$. If no solution is found, one increases $r$ until a solution 
is found.
The practical limitation of our method is the ability to solve large systems of linear equations.

The method presented here grew out of our previous work on the two-loop sunrise integral with unequal masses in two space-time dimensions.
In this work we used methods of algebraic geometry and derived an ordinary differential equation for this integral.
For the differential equation we have to find the differential operator appearing on the left-hand side, as well as the inhomogeneous term
appearing on the right-hand side.
In our previous work this was done in a two-step process: We first obtained the differential operator, and then determined the inhomogeneous term
by an ansatz.
To find the differential operator we used the fact that for this specific integral the singularities of the integrand correspond to an elliptic curve.
We then computed the differential operator from the known cohomology of the elliptic curve.

For an arbitrary Feynman integral the singularities of the integrand correspond no longer to elliptic curves and generalise to algebraic 
varieties. The cohomology of these algebraic varieties is in general not known a priori.
The approach of our previous work does therefore not generalise trivially to arbitrary Feynman integrals.
However, a careful inspection shows that we can encode all required unknown information for the differential operator
and the inhomogeneous term into coefficients of a well-chosen ansatz.
Plugging the ansatz into the differential equation will give a system of linear equations for the unknown coefficients.
We thus reduce the problem of finding the differential equation to the (simpler) problem of solving a system of linear equations.

In addition, our approach has a few bonuses, which we would like to point out:
First of all, whereas methods of algebraic geometry are often tied to fixed integer dimensions, nothing prevents us from
using an arbitrary space-time dimension within our ansatz. We are thus able to derive a differential equation within dimensional 
regularisation.
Secondly, when working within dimensional regularisation, we observe interesting factorisation properties of the differential operator.
These factorisation properties are useful when the differential equation is solved term by term as a Laurent series in the dimensional
regularisation parameter.
Thirdly, our ansatz leads to a straightforward interpretation of the inhomogeneous term: It can be represented as a sum
of Feynman integrals, where one of the original propagators has been removed.
These Feynman integrals can be considered as simpler Feynman integrals.

It is worth mentioning that (co-)homological methods have been used in connection with Feynman integrals already a long time ago \cite{Hwa,Boyling:1968,Golubeva:1970gr,Lefschetz}.
Renewed interest in this topic has grown out of the seminal paper by Bloch, Esnault and Kreimer \cite{Bloch:2006}.

This paper is organised as follows:
In the next section we introduce our notation.
Section~\ref{sect:method} presents the method for finding the differential equation.
A few examples are discussed in section~\ref{sect:examples}.
Finally, section~\ref{sect:conclusions} contains our conclusions.

\section{Notation and overview}
\label{sect:notation}

We start by defining our notation.
We consider a $l$-loop Feynman graph with $m$ external lines and $n$ internal lines.
We label the external momenta by $p_1$, ..., $p_m$ and the $l$ independent loop momenta by
$k_1$, ..., $k_l$. The momenta flowing through the internal lines are denoted by $q_i$ with $1 \le i \le n$.
The momenta $q_i$ can be expressed as a linear combination of the external momenta $p_j$ 
and the independent loop momenta $k_j$ with coefficients $-1$, $0$ or $1$:
\bq
\label{eq_internal_mom}
 q_i & = & \sum\limits_{j=1}^l \rho_{ij} k_j + \sum\limits_{j=1}^m \sigma_{ij} p_j,
 \;\;\;\;\;\; 
 \rho_{ij}, \sigma_{ij} \in \{-1,0,1\}.
\eq
The Lorentz invariant quantities in the external momenta are
\bq
 s_{jk} & = & \left( p_j + p_k \right)^2, 
 \;\;\; 1 \le j,k \le m.
\eq
The internal masses are denoted by $m_i$ with $1 \le i \le n$.
We call the set
\bq
\label{def_set_invariants}
 S & = & \left\{ s_{jk}, m_i^2 \right\}
\eq
the set of kinematical invariants.
We denote by $D \in {\mathbb C}$ the dimension of space-time. 
Within dimensional regularisation $D$ is extended from positive integer values to complex values.
For each internal line the number $\nu_i \in {\mathbb N}$ (with $1 \le i \le n$) gives the power to which the $i$-th
propagator occurs.
We define the Feynman integral by
\bq
\label{eq0}
I_G  & = &
 \frac{\prod\limits_{j=1}^{n}\Gamma(\nu_j)}{\Gamma(\nu-lD/2)}
 \left( \mu^2 \right)^{\nu-l D/2}
 \int \prod\limits_{r=1}^{l} \frac{d^Dk_r}{i\pi^{\frac{D}{2}}}\;
 \prod\limits_{j=1}^{n} \frac{1}{(-q_j^2+m_j^2)^{\nu_j}},
\eq
with $\nu=\nu_1+...+\nu_n$.
The scale $\mu$ is an arbitrary parameter with the dimension of a mass, which is introduced to make the integral dimensionless.
After Feynman parametrisation we have
\bq
\label{eq1}
I_G  & = &
 \int\limits_{\Delta}  f \; \omega.
\eq
Eq.~(\ref{eq1}) motivates the choice of the prefactor in eq.~(\ref{eq0}): The prefactor is chosen such that after Feynman parametrisation
one obtains a nice formula. 
The integration in eq.~(\ref{eq1}) is over
\bq
 \Delta & = & \left\{ \left[ x_1 : x_2 : ... : x_n \right] \in {\mathbb P}^{n-1} | x_i \ge 0, 1 \le i \le n \right\}.
\eq
$\omega$ is a differential $(n-1)$-form given by
\bq
 \omega & = & \sum\limits_{j=1}^n (-1)^{j-1}
  \; x_j \; dx_1 \wedge ... \wedge \widehat{dx_j} \wedge ... \wedge dx_n,
\eq
where the hat indicates that the corresponding term is omitted.
The function $f$ is given by
\bq
f(x_1,...,x_n) & = &
 \left( \mu^2 \right)^{\nu-l D/2}
 \left( \prod\limits_{j=1}^n x_j^{\nu_j-1} \right)
 \frac{{\mathcal U}^{\nu-(l+1) D/2}}{{\mathcal F}^{\nu-l D/2}}.
\eq
The functions ${\mathcal U}$ and ${\mathcal F}$ are the graph polynomials.
Graph polynomials have a long history dating back to Kirchhoff and there  
are well-established books and reviews on this subject \cite{Eden,Nakanishi,Todorov,Zavialov,Itzykson:1980rh,Smirnov:2006ry,Bogner:2010kv}. 
If one expresses
\bq
 \sum\limits_{j=1}^{n} x_{j} (-q_j^2+m_j^2)
 & = & 
 - \sum\limits_{r=1}^{l} \sum\limits_{s=1}^{l} k_r M_{rs} k_s + \sum\limits_{r=1}^{l} 2 k_r \cdot Q_r - J,
\eq
where $M$ is a $l \times l$ matrix with scalar entries and $Q$ is a $l$-vector
with four-vectors as entries,
one obtains
\bq
 {\mathcal U} = \mbox{det}(M),
 & &
 {\mathcal F} = \mbox{det}(M) \left( - J + Q M^{-1} Q \right).
\eq 
The graph polynomials ${\mathcal U}$ and ${\mathcal F}$ 
are homogeneous functions in the Feynman parameters $x_j$ with degrees
\bq
\mbox{deg}\;{\mathcal U} = l,
 & &
\mbox{deg}\;{\mathcal F} = l+1.
\eq
The graph polynomial ${\cal U}$ is independent of the kinematical invariants, while the graph polynomial ${\cal F}$ depends linearly on the
kinematical invariants.
Let us denote by $t \in S$ one kinematical invariant.
$t$ will be our main variable in the following.
We seek an ordinary differential equation with respect to the variable $t$ for the Feynman integral $I_G$.
We set
\bq
 \omega_t & = & 
 f \omega,
\eq
and hence
\bq
 I_G & = & \int\limits_\Delta \omega_t.
\eq
$\omega_t$ is a $(n-1)$-form which is homogeneous of degree $-(n-1)$ in the variables $x_i$.
We look for a differential equation of the form
\bq
\label{basic_picard}
 L^{(r)} \omega_t & = & d \beta,
\eq
where
\bq
\label{picard_fuchs_operator}
 L^{(r)} & = & \sum\limits_{j=0}^r p_j \left( \mu^2 \frac{d}{dt} \right)^j
\eq
is a Picard-Fuchs operator of order $r$.
The coefficients $p_j$ may depend on the kinematical invariants from the set $S$, the scale $\mu^2$, the space-time
dimension $D$ and the exponents $\nu_i$, but not on the Feynman parameters $x_i$.
We normalise the Picard-Fuchs operator such that $p_r=1$.
$\beta$ is a $(n-2)$-form, depending on the Feynman parameters $x_i$.
The differential $d$ is with respect to the Feynman parameters $x_i$.
The differential $(n-2)$-form $\beta$ is not unique.
We may add to $\beta$ any closed $(n-2)$-form $\gamma$ without changing eq.~(\ref{basic_picard}).

Suppose an equation of the form as in eq.~(\ref{basic_picard}) exists.
Integration yields then
\bq
 L^{(r)} I_G & = & \int\limits_\Delta d \beta.
\eq
If the conditions for Stokes' theorem are met, we obtain
\bq
\label{differential_equation}
 L^{(r)} I_G & = & \int\limits_{\partial \Delta} \beta.
\eq
Within dimensional regularisation there are no obstructions in using Stokes' theorem. If our method is used in fixed space-time
dimensions, say $D=4$, it may happen that although the left-hand side is finite, the individual sub-integrals over the faces of $\Delta$ are
divergent. In this case one replaces ${\mathbb P}^{n-1}$ by a suitable blow-up $P$ and applies Stokes' theorem to the integration in $P$.

Eq.~(\ref{differential_equation}) is the sought-after ordinary differential equation for the Feynman integral $I_G$.
$L^{(r)}$ is a differential operator of order $r$ in the variable $t$. 
The right-hand side is given as a sum of integrals with $(n-1)$ Feynman parameters.
These integrals correspond to graphs with one propagator less and can be considered as simpler.
If the ``simpler'' integrals are known (the right-hand side and the boundary values, say at $t=0$),
then eq.~(\ref{differential_equation}) can be used to obtain the Feynman integral $I_G$.
This requires to solve the differential equation in eq.~(\ref{differential_equation}).
But first, of course, one has to find the differential equation.
We show that the problem of finding the differential equation can be reduced to solving linear equations.

We comment on our use of the word ``Picard-Fuchs''.
A differential operator of the form as in eq.~(\ref{picard_fuchs_operator}) is said to be of Fuchsian type, if all coefficients $p_j$ are
meromorphic functions of $t$ and if $p_j$ has at most poles of order $(r-j)$.
Within our method, once the differential equation has been found, the Fuchsian property is easily verified a posteriori.
The terminus ``Picard-Fuchs'' is usually associated with the following situation: Consider a family of algebraic varieties
depending on a parameter $t$ and a cohomology class in $H^{n-1}$. One is interested in the variation of the cohomology with the parameter $t$.
In order to pass from cohomology classes to functions, one considers periods of the cohomology classes.
Then the variation of the periods with the parameter $t$ is described by a differential equation of Fuchsian type.
This equation is refered to as a Picard-Fuchs equation.
The relation to Feynman integrals is as follows: Fix an even integer space-time dimension and suppose that the Feynman integral is finite in this
space-time dimension.
For even integer space-time dimension the integrand in eq.~(\ref{eq1}) is a rational form on ${\mathbb P}^{n-1}$ with possible singularities
on the zero sets of ${\mathcal F}$ and ${\mathcal U}$. In order to resolve the singularities one considers a blow-up $P$ of ${\mathbb P}^{n-1}$
and one denotes by $Y$ the strict transform of the variety on which the integrand is singular. One further denotes by $B$ the total transform
of the integration boundary in eq.~(\ref{eq1}). According to \cite{Bloch:2006} the integrand corresponds to a (relative) cohomology class of
\bq
 H^{n-1}\left(P\backslash Y, B \backslash B \cap Y \right)
\eq
and the Feynman integral to a period of this cohomology class.
The variation of this period with the parameter $t$ is given by the differential equation in eq.~(\ref{differential_equation}).
In this way eq.~(\ref{differential_equation}) is a Picard-Fuchs equation.
Now eq.~(\ref{differential_equation}) still makes sense if we relax the assumption on even integer space-time dimensions and on the finiteness
of the Feynman integral. We can consider divergent Feynman integrals regulated by dimensional regularisation.
Also in this case we find a differential equation of the form as in eq.~(\ref{differential_equation}).
We continue to call this equation a Picard-Fuchs equation.

\section{The method for finding the differential equation}
\label{sect:method}

The method for finding the differential equation can be summarised as follows:
In the beginning the order $r$ of the differential equation, the coefficients $p_j$ ($0 \le j < r$) and the differential form $\beta$ are unknown.
We are interested in finding the lowest-order differential equation.
To this aim we start with $r=1$ and an ansatz for the coefficients $p_j$ and the differential form $\beta$.
From this ansatz we obtain a system of linear equations. If a solution for this system exists we are done.
Otherwise we repeat the game with $r+1$.
At least for finite integrals in even integer space-time dimensions it is guaranteed that 
this ansatz works since systems of differential equations coming from algebraic geometry are known to
have regular singular points.

Our ansatz is based on the fact, that the singularities of the integrand are given by powers of the graph polynomials ${\cal U}$ and ${\cal F}$.
In other words, all singularities lie in Feynman parameter space ${\mathbb P}^{n-1}$ on the hyper-surfaces defined by
\bq
 {\cal U} = 0,
 & &
 {\cal F} = 0.
\eq
Polynomials in the Feynman parameters in the numerator may cancel some of the singularities, 
but the main point is that there will be no singularities on new algebraic varieties.
This is obvious on the left-hand side: Acting with $L^{(r)}$ on the integrand will only increase the power of ${\cal F}$ in the denominator by $r$.
This motivates the following ansatz for the differential $(n-2)$-form $\beta$:
\bq
\label{ansatz_beta}
 \beta & = & 
 \left( \mu^2 \right)^{r-1}
 \frac{f}{{\cal F}^{r-1}} \alpha,
\eq
where $\alpha$ is again a $(n-2)$-form, but free of singularities.

Secondly we exploit the fact that $I_G$ is an integral on projective space ${\mathbb P}^{n-1}$.
This implies that the $(n-2)$-form $\alpha$ must be homogeneous of degree $[(r-1)(l+1)+2]$ in the variables $x_i$.
The degree of homogeneity of $\alpha$ is always a positive integer.
From this and the fact that $\alpha$ is free of singularities we are led to the ansatz that $\alpha$ can be represented by polynomials
of degree $[(r-1)(l+1)+2]$
in the Feynman parameters $x_j$.

Following Griffiths \cite{Griffiths:1969} 
we can reduce the degree of the unknown polynomials by one by assuming that $\alpha$ can be written in the form
\bq
\label{ansatz_alpha}
 \alpha & = & 
 \sum\limits_{j_1<j_2} \left(-1\right)^{j_1+j_2}
 \left[ -x_{j_1} a_{j_2} + x_{j_2} a_{j_1} \right]
 dx_1 \wedge ... \wedge \widehat{dx_{j_1}} \wedge ... \wedge \widehat{dx_{j_2}} \wedge ... \wedge dx_n,
\eq
where the $a_i$ are homogeneous polynomials of degree $h=[(r-1)(l+1)+1]$ in the variables $x_i$.
For the polynomials $a_i$ we assume the most general form.
For example, if $n=3$ and $h=2$ we have
\bq
\label{example_expansion_monomials}
 a_i
 & = & 
 a^{(i)}_{200} x_1^2 + a^{(i)}_{020} x_2^2 + a^{(i)}_{002} x_3^2
 + a^{(i)}_{110} x_1 x_2 + a^{(i)}_{011} x_2 x_3 + a^{(i)}_{101} x_1 x_3.
\eq
The variables $a^{(i)}_{jkl}$ are independent of the Feynman parameters.
The most general homogeneous polynomial of degree $h$ in $n$ variables has
\bq
 \left( \begin{array}{c}
 n+h-1 \\
 h \\
 \end{array} \right)
\eq
monomials.
This is the number of possibilities to pick $h$ elements out of a set of $n$ elements by not taking the order into account and with repetitions.
For a given $r$ we therefore have within our ansatz as unknown variables the variables $p_0$, $p_1$, ..., $p_{r-1}$, which appear in the
Picard-Fuchs operator (recall that our convention is to take $p_r=1$), as well as all coefficients appearing in the expansion into monomials
of the polynomials $a_i$. In the example of eq.~(\ref{example_expansion_monomials}) these are the variables $a^{(i)}_{jkl}$.
In general, we have
\bq
\label{number_unknowns}
 N_{\mathrm{unknowns}}
 & = &
 r + n
 \left( \begin{array}{c}
 (r-1)(l+1)+n \\
 (r-1)(l+1)+1 \\
 \end{array} \right)
\eq
unknown variables. The factor $n$ in front of the binomial coefficient comes from the fact that we have $n$ polynomials $a_i$.

Now let us plug our ansatz, specified by eq.~(\ref{ansatz_beta}) and eq.~(\ref{ansatz_alpha}) into eq.~(\ref{basic_picard}).
Working out $d\beta$ we find
\bq
 d\beta & = &
 \left( \mu^2 \right)^{r-1}
 \sum\limits_{j=1}^n \left(-1\right)^{j-1}
 \sum\limits_{i=1}^n
 \partial_i \left[  \frac{f}{{\cal F}^{r-1}} \left( -x_i a_j + x_j a_i \right) \right]
 dx_1 \wedge ... \wedge \widehat{dx_{j}} \wedge ... \wedge dx_n.
\eq
Plugging this expression into eq.~(\ref{basic_picard}) and comparing the coefficients of 
$dx_1 \wedge ... \wedge \widehat{dx_{j}} \wedge ... \wedge dx_n$ we obtain 
for $1 \le j \le n$ the equations
\bq
\label{equation_j}
 x_j 
 L^{(r)} f 
 & = & 
 \left( \mu^2 \right)^{r-1} 
 \sum\limits_{i=1}^n
 \partial_i \left[  \frac{f}{{\cal F}^{r-1}} \left( -x_i a_j + x_j a_i \right) \right].
\eq
We multiply both sides in eq.~(\ref{equation_j}) by
\bq
 \left(\mu^2\right)^{-\nu+l D/2-r}
 {\cal F}^{\nu-l D/2 + r} {\cal U}^{-\nu + (l+1) D/2 + 1} 
 \left( \prod\limits_{j=1}^n x_j^{-\nu_j+2} \right).
\eq
This factor is chosen such that after multiplication both sides of the equation are 
homogeneous polynomials in the variables $x_i$.
We obtain
\bq
\label{master_equation}
 0 & = &
 x_j \left( \prod\limits_{k=1}^n x_k \right) {\cal U}
 \sum\limits_{s=0}^r 
 p_s
 \frac{\Gamma\left(l\frac{D}{2} -\nu+1\right)}{\Gamma\left(l\frac{D}{2} -\nu+1-s\right)}
 \left(\mu^2\right)^{s-r}
 {\cal F}^{r-s}
 \dot{\cal F}^s
 \nonumber \\
 & & 
 - \sum\limits_{i=1}^n 
 \left\{
 \left( \prod\limits_{k=1}^n x_k \right) 
   {\cal U} \frac{{\cal F}}{\mu^2}
 \partial_i \left( -x_i a_j + x_j a_i \right)
 + 
 \left(\nu_i-1 \right)
 \left( \prod\limits_{k=1,k\neq i}^n x_k \right) 
   {\cal U} \frac{{\cal F}}{\mu^2} \left( -x_i a_j + x_j a_i \right)
 \right. \nonumber \\
 & & \left.
 + 
 \left(\nu - \left(l+1\right)\frac{D}{2} \right)
 \left( \prod\limits_{k=1}^n x_k \right) 
   \left( \partial_i {\cal U} \right) \frac{{\cal F}}{\mu^2} \left( -x_i a_j + x_j a_i \right)
 \right. \nonumber \\
 & & \left.
 + 
 \left(l \frac{D}{2} - \nu -r + 1 \right)
 \left( \prod\limits_{k=1}^n x_k \right) 
   {\cal U} \frac{\left( \partial_i {\cal F}\right) }{\mu^2} \left( -x_i a_j + x_j a_i \right)
 \right\}.
\eq
In eq.~(\ref{master_equation}) we denoted the derivative with respect to $t$ by
\bq
 \dot{\cal{F}} & = & \frac{d}{dt} {\cal F}.
\eq
Eq.~(\ref{master_equation}) is our master equation and we should pause a moment to contemplate the main features of this equation.
First of all, each term of this equation is of degree one or zero in the 
unknown variables (the coefficients $p_j$ and the coefficients appearing in the expansion of the polynomials $a_i$ into
monomials).
Secondly, eq.~(\ref{master_equation}) is homogeneous of degree $[n+(l+1)(r+1)]$ in the variables $x_i$.
Since eq.~(\ref{master_equation}) has to hold for all values of the variables $x_i$, the coefficient $c$ of each monomial in the variables $x_i$ has
to vanish.
But each coefficient $c$ of a monomial in the variables $x_i$ yields a linear equation $c=0$ in the unknown variables.
We thus obtain a (possibly large) system of linear equations for the unknown variables.
In total we obtain by this method 
\bq
\label{number_equations}
 N_{\mathrm{equations}}
 & = &
 n
 \left( \begin{array}{c}
 (l+1)(r+1)+2n-1 \\
 (l+1)(r+1)+n \\
 \end{array} \right)
\eq
equations for $N_{\mathrm{unknowns}}$. The number $N_{\mathrm{unknowns}}$ has been given in eq.~(\ref{number_unknowns}).
Of course, not all equations will be independent.
With methods from linear algebra we may attempt to solve this system. 
If the system admits a solution, we have found a differential equation for the Feynman integral under consideration.
In the case where a solution exists, there will be in general more than one solution.
This is related to the fact that one can always add a closed $(n-2)$-form to $\beta$ in eq.~(\ref{basic_picard}).
This does not affect our method. In order to find the differential equation one can pick any solution.

If the system does not admit a solution, we repeat the exercise by increasing the order $r$ of the differential operator $L^{(r)}$ by one.
In general this will result in a linear system with more unknowns and more equations.
The practical limitation of this method is the ability to solve large systems of linear equations.

We illustrate our method with a simple example. Consider the integral
\bq
 I & = &
 \mu^2
 \int\limits_{x_j \ge 0} d^{2}x \; \delta\left(1-x_1-x_2\right) \;
 \frac{1}{{\mathcal F}},
 \;\;\;\;\;\;
 {\mathcal F} = -t x_1 x_2 + \left(x_1+x_2\right)^2 m^2.
\eq
This corresponds to a one-loop two-point function with equal internal masses in $D=2$ space-time dimensions.
Suppose we search for a first-order Picard-Fuchs operator
\bq
 L^{(1)} & = & \mu^2 \frac{d}{dt} + p_0.
\eq
For $\beta$ our ansatz reads
\bq
 \beta & = &
 \frac{\mu^2}{{\mathcal F}} \left( x_1 a_2 - x_2 a_1 \right),
\eq
where $a_1$ and $a_2$ are homogeneous polynomials of degree $1$ in the variables $x_1$ and $x_2$. We write
\bq
 a_1 & = & a^{(1)}_{10} x_1 + a^{(1)}_{01} x_2,
 \nonumber \\
 a_2 & = & a^{(2)}_{10} x_1 + a^{(2)}_{01} x_2.
\eq
We have two Feynman parameters and therefore we have to consider the cases $j=1$ and $j=2$ in eq.~(\ref{master_equation}).
In this example eq.~(\ref{master_equation}) is homogeneous of degree 6. The coefficients of the monomials in the variables $x_i$ must vanish.
This yields the equations
\bq
 m^2 p_0 + m^2 a^{(1)}_{10} + \left(2m^2-t\right) a^{(2)}_{10} - m^2 a^{(2)}_{01} & = & 0,
 \nonumber \\
 m^2 p_0 -m^2 a^{(1)}_{10} + \left(2m^2-t\right) a^{(1)}_{01} + m^2 a^{(2)}_{01} & = & 0,
 \nonumber \\
 \left(3m^2-t\right) p_0 - m^2 a^{(1)}_{10} + \left(4m^2-t\right) a^{(1)}_{01} + 2 m^2 a^{(2)}_{10} +m^2 a^{(2)}_{01} & = & -\mu^2,
 \nonumber \\
 \left(3m^2-t\right) p_0 + m^2 a^{(1)}_{10} + 2 m^2 a^{(1)}_{01} + \left(4m^2-t\right) a^{(2)}_{10} -m^2 a^{(2)}_{01} & = & -\mu^2.
\eq
This is a system of linear equations in the unknown variables $\{p_0, a^{(1)}_{10}, a^{(1)}_{01}, a^{(2)}_{10}, a^{(2)}_{01} \}$.
In order to determine the differential equation we are only interested if the system admits a solution, and if this is the case in one particular
solution.
The system of linear equations for this example is easily solved.
One finds
\bq
 p_0 & = & \frac{\mu^2\left(t-2m^2\right)}{t\left(t-4m^2\right)},
 \nonumber \\
 a^{(1)}_{01} & = & \frac{2\mu^2 m^2-t\left(t-4m^2\right) a^{(2)}_{10}}{t\left(t-4m^2\right)}
 \nonumber \\
 a^{(2)}_{01} & = & \frac{\mu^2 m^2\left(t-2m^2\right) +m^2 t \left(t-4m^2\right) a^{(1)}_{10} - t \left(t-2m^2\right)\left(t-4m^2\right) a^{(2)}_{10}}{m^2 t\left(t-4m^2\right)}.
\eq
The solutions are parametrised by $a^{(1)}_{10}$ and $a^{(2)}_{10}$. We can choose any values for these two variables. 
In particular we can set $a^{(1)}_{10}=a^{(2)}_{10}=0$.
This freedom of choice corresponds to the fact that we can always add a closed form to $\beta$.

Let us have a closer look at the inhomogeneous term
\bq
 \int\limits_{\partial \Delta} \beta
\eq
in eq.~(\ref{differential_equation}). Within dimensional regularisation the integration is over the $n$ faces of the simplex $\Delta$.
Let us consider what happens on a particular face. Without loss of generality we can consider the $n$-th face.
We consider the restriction of $\beta$ to $x_n=0$.
If $\nu_n > 1$ we have
\bq
 \left. \beta \right|_{x_n=0} & = & 0.
\eq
Otherwise, if $\nu_n=1$ we find
\bq
 \left. \beta \right|_{x_n=0} & = & 
 \left(-1\right)^n 
 \left( \mu^2 \right)^{\nu-l D/2+r-1}
 a_n
 \left( \prod\limits_{i=1}^{n-1} x_i^{\nu_i-1} \right)
 \left. \frac{{\mathcal U}^{\nu-(l+1) D/2}}{{\mathcal F}^{\nu-l D/2+r-1}} \right|_{x_n=0}
 \nonumber \\
 & & 
 \times
 \sum\limits_{j=1}^{n-1} (-1)^{j-1}
  \; x_j \; dx_1 \wedge ... \wedge \widehat{dx_j} \wedge ... \wedge dx_{n-1}.
\eq
Now let
\bq
 a_n & = &
 \sum\limits_{\substack{m_1\ge0,..,m_n\ge0\\m_1+...+m_n=(r-1)(l+1)+1}} a^{(n)}_{m_1 ... m_n} x^{m_1} ... x^{m_n}
\eq
be the expansion of the polynomial $a_n$ into monomials. We recall that $a_n$ is homogeneous of degree $(r-1)(l+1)+1$.
On the $n$-th face only the monomials with $m_n=0$ are relevant.
We thus have
\bq
 \left. \beta \right|_{x_n=0} & = & 
 \left(-1\right)^n 
 \sum\limits_{\substack{m_1\ge0,..,m_{n-1}\ge0\\m_1+...+m_{n-1}=(r-1)(l+1)+1}} a^{(n)}_{m_1 ... m_{n-1}0}
 \left( f^{(n)}_{m_1 ... m_{n-1}} \omega^{(n)} \right),
\eq
with
\bq
 f^{(n)}_{m_1 ... m_{n-1}} & = &
 \left( \mu^2 \right)^{\nu-l D/2+r-1}
 \left( \prod\limits_{i=1}^{n-1} x_i^{\nu_i+m_i-1} \right)
 \left. \frac{{\mathcal U}^{\nu-(l+1) D/2}}{{\mathcal F}^{\nu-l D/2+r-1}} \right|_{x_n=0},
 \nonumber \\
 \omega^{(n)} & = &
 \sum\limits_{j=1}^{n-1} (-1)^{j-1}
  \; x_j \; dx_1 \wedge ... \wedge \widehat{dx_j} \wedge ... \wedge dx_{n-1}.
\eq
We remind the reader that
\bq
 \left.{\mathcal U}\right|_{x_n=0}
 & \mbox{and} &
 \left.{\mathcal F}\right|_{x_n=0}
\eq
are the graph polynomials of a graph, obtained from the original one by contracting the edge $n$ \cite{Bogner:2010kv}.
Therefore
\bq
 f^{(n)}_{m_1 ... m_{n-1}} \omega^{(n)}
\eq
is the integrand of a Feynman integral in $(D+2r-2)$ space-time dimensions, where the propagator $n$ has been contracted.
Thus the inhomogeneous term in eq.~(\ref{differential_equation}) is given as a linear combination of Feynman integrals in $(D+2r-2)$ space-time
dimensions, where one of the $n$ propagators of the original integral has been contracted.
We remark that the representation of the inhomogeneous term as a linear combination of Feynman integrals is not necessarily unique.
As previously already mentioned,
we can can always add a linear combination of Feynman integrals corresponding to a closed form $\gamma$.

\section{Examples}
\label{sect:examples}

In this section we discuss a few examples.
Since the size of the system of linear equations which needs to be solved grows quite fast with order of the Picard-Fuchs operator, the number of
loops and the number of propagators, we limit us here to examples which can be solved with plain vanilla methods within standard computer algebra
systems.
A dedicated and more efficient method to solve these systems of linear equations is work in progress.

As already mentioned, the inhomogeneous term is not unique. We can always add a closed form to $\beta$, which corresponds to adding a linear combination
of (simpler) integrals to the inhomogeneous term, which adds up to zero.
As a consequence, the inhomogeneous term usually does not have a particular ``nice'' form and it is usually rather lengthy.
On the other hand, the solution of the linear system for the Picard-Fuchs operator is unique and usually rather compact.
For this reason we present here only the Picard-Fuchs operators for several examples.
But the reader should keep in mind that by solving the system of linear equations one obtains simultaneously the solutions for the inhomogeneous terms.

In particular we would like to give the reader with the chosen examples some indications, what order can be expected for the Picard-Fuchs operator.
As examples we study the family of ``banana'' graphs.
The $l$-loop banana graph is a two-point functions with $(l+1)$ propagators.
\begin{figure}
\begin{center}
\begin{picture}(100,100)(0,0)
\Vertex(20,50){2}
\Vertex(80,50){2}
\Line(80,50)(100,50)
\Line(0,50)(20,50)
\CArc(50,50)(30,0,180)
\CArc(50,50)(30,180,360)
\CArc(50,20)(42.43,45,135)
\CArc(50,80)(42.43,225,315)
\Text(105,50)[l]{$p$}
\end{picture}
\end{center}
\caption{
The three-loop banana graph.
}
\label{fig_banana_graph}
\end{figure}
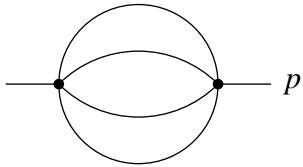
As an example, the three-loop banana graph is shown in fig.~(\ref{fig_banana_graph}).
The $l$-loop banana graph is defined by
\bq
I_l  & = &
 \frac{\left( \mu^2 \right)^{l+1-l D/2}}{\Gamma(l+1-l D/2)}
 \int 
 \left( \prod\limits_{r=1}^{l} \frac{d^Dk_r}{i\pi^{\frac{D}{2}}}\; \frac{1}{(-k_r^2+m_r^2)} \right)
 \frac{1}{\left[-\left(p-k_1-...-k_l\right)^2+m_{l+1}^2\right]}
 \nonumber \\
 & = &
 \left( \mu^2 \right)^{l+1-l D/2} 
 \int\limits_{x_j \ge 0} d^{l+1}x \; \delta\left(1-\sum\limits_{j=1}^{l+1} x_j \right) \;
 \frac{{\mathcal U}_l^{(l+1) (1-D/2)}}{{\mathcal F}_l^{l+1-l D/2}},
\eq
with
\bq
 {\mathcal U}_l = \left( \prod\limits_{i=1}^{l+1} x_i \right) \left( \sum\limits_{j=1}^{l+1} \frac{1}{x_j} \right),
 & & 
 {\mathcal F}_l = - t \left( \prod\limits_{i=1}^{l+1} x_i \right) + {\mathcal U} \left( \sum\limits_{j=1}^{l+1} x_j m_j^2 \right)
\eq
and $t=p^2$.
In $D=2$ dimensions these integral simplify to
\bq
 I_l & = &
 \mu^2
 \int\limits_{x_j \ge 0} d^{l+1}x \; \delta\left(1-\sum\limits_{j=1}^{l+1} x_j \right) \;
 \frac{1}{{\mathcal F}_l},
\eq
and depend only on the graph polynomial ${\mathcal F}$, but not on the graph polynomial ${\mathcal U}$.
In addition we consider first the equal mass case $m_1=m_2...=m_{l+1}=m$.
In the equal mass case and in $D=2$ dimensions the Picard-Fuchs operator are particular simple.
At one-loop ($l=1$) we find that 
the Picard-Fuchs operator is of order $1$ and given by
\bq
\label{oneloop_D2_equalmass}
 L^{(1)} & = & 
 \mu^2 
 \left[ \frac{d}{dt} 
 + \frac{t-2m^2}{t\left(t-4m^2\right)} \right].
\eq
This example was discussed in detail in the previous section.
For $l=2$ the Picard-Fuchs operator is of order $2$ and reads
\bq
\label{twoloop_D2_equalmass}
 L^{(2)} & = &
 \left( \mu^2 \right)^2
 \left[
 \frac{d^2}{dt^2} + \frac{3t^2 - 20 m^2 t + 9 m^4}{t\left(t-m^2\right)\left(t-9m^2\right)} \frac{d}{dt}
 + \frac{t-3m^2}{t\left(t-m^2\right)\left(t-9m^2\right)}
 \right].
\eq
The results for $l=1$ and $l=2$ have been known for a long time. With our method we obtain also the Picard-Fuchs operator
for the three-loop diagram:
\bq
 L^{(3)} = 
 \left( \mu^2 \right)^3
 \left[
 \frac{d^3}{dt^3}
 + \frac{6\left(t^2-15m^2t+32m^4\right)}{t \left( t-4m^2 \right) \left(t-16m^2\right)} \frac{d^2}{dt^2}
 + \frac{7t^2 -68 m^2 t + 64 m^4}{t^2 \left( t-4m^2 \right) \left(t-16m^2\right)}\frac{d}{dt}
 + \frac{1}{t^2 \left(t-16m^2\right)}
 \right].
 \nonumber
\eq
In $D=2$ space-time dimensions the
order of the Picard-Fuchs operator for the family of the banana graphs is independent of the mass parameters, as long as the masses are non-zero. 
This is related to the fact, that in this case the singularities of the integrand are only determined by the graph polynomial ${\mathcal F}$, but
not by ${\mathcal U}$.
In the unequal mass case
eq.~(\ref{oneloop_D2_equalmass}) generalises to
\bq
\label{oneloop_D2_unequalmass}
 L^{(1)} & = & 
 \mu^2 
 \left[ \frac{d}{dt} 
 + \frac{t-m_1^2-m_2^2}{\left[t-\left(m_1-m_2\right)^2\right]\left[t-\left(m_1+m_2\right)^2\right]} \right].
\eq
The generalisation towards arbitrary masses of eq.~(\ref{twoloop_D2_equalmass}) has been given in \cite{MullerStach:2011ru}.

Our method can also be used to derive the differential equation in $D$ dimensions. 
For the one-loop case we obtain in $D$ dimensions
\bq
 L^{(1)} & = & 
 \mu^2 
 \left[ \frac{d}{dt} 
 + \frac{\left(4-D\right)t^2-2\left(m_1^2+m_2^2\right)t+\left(D-2\right)\left(m_1^2-m_2^2\right)^2}{2t\left[t-\left(m_1-m_2\right)^2\right]\left[t-\left(m_1+m_2\right)^2\right]} \right],
\eq
In the two-loop case with unequal masses we find in $D$ dimensions a Picard-Fuchs operator of order $4$.
The increase in the order of the Picard-Fuchs operator compared to the $D=2$ case is explained by the fact, that away from $D=2$ space-time dimensions
also the graph polynomial ${\mathcal U}$ can contribute to the singularities of the integrand.
However, we observe that this fourth-order differential operator factorises in four space-time dimensions into a differential operator of order two
and two differential operators of order one.
We illustrate this first with a simple example: Let us consider the integral
\bq
 \tilde{I}_{1} & = & 
 \mu^2 \int\limits_{x_j \ge 0} d^2x \; \delta\left(1-x_1-x_2\right) \; \frac{\left(x_1-x_2\right)}{{\cal U}_1{\cal F}_1}.
\eq
This integral corresponds to a one-loop integral in $D=4$ space-time dimensions, where one propagator occurs to the power two.
Unlike $I_1$ in eq.~(\ref{oneloop_D2_unequalmass}), the integral $\tilde{I}_{1}$ has a second order Picard-Fuchs operator, which
can be written in factorised form as 
\bq
\lefteqn{
 L^{(2)} = 
 } & & 
 \\
 & & 
 \left(\mu^2\right)^2
 \left\{ \frac{d}{dt} 
 + \frac{\left[ 3 t^3 - 7 \left(m_1^2+m_2^2\right)t^2 + 5 \left(m_1^2+m_2^2\right)^2 t
          - \left(m_1^2+m_2^2\right)\left(m_1^2-m_2^2\right)^2 \right]}{t\left[t-m_1^2-m_2^2\right]\left[t-\left(m_1-m_2\right)^2\right]\left[t-\left(m_1+m_2\right)^2\right]}
 \right\}
 \left[ \frac{d}{dt} + \frac{1}{t} \right].
 \nonumber 
\eq
Note that the second factor is independent of the masses.

Turning back to the two-loop graph with unequal masses in $D$ dimensions we can write the fourth-order differential operator $L^{(4)}$ for this integral
in $D$ dimensions as
\bq
\label{factorisation}
 L^{(4)} & = &
 L^{(2)}_{\mathrm{reduced}} 
 \left(\mu^2\right)^2
 \left( \frac{d}{dt} + \frac{2}{t} \right) \frac{d}{dt} + \frac{\left( 4 - D \right)}{2} L^{(4)}_{\eps},
\eq
where $L^{(2)}_{\mathrm{reduced}}$ and $L^{(4)}_{\eps}$ are differential operators of order $2$ and $4$, respectively.
Writing out these two operators leads to rather long expressions, but the explicit expression is not needed for the discussion here.
$L^{(2)}_{\mathrm{reduced}}$ is independent of the dimensional regularisation parameter $\eps=(4-D)/2$, while $L^{(4)}_{\eps}$ has a Taylor expansion
in the parameter $\eps$.
The factorisation property in eq.~(\ref{factorisation}) is very helpful in solving the differential equation:
If the differential equation is solved term by term as a Laurent series in the 
dimensional regularisation parameter $\eps$, the term with the operator $L^{(4)}_{\eps}$ has an extra factor of $\eps$ in front of it and can thus be
treated as part of the inhomogeneous term.
This reduces the problem of solving a fourth-order differential equation to solving a 
second-order differential equations plus two integrations corresponding to the two linear factors.

\section{Conclusions}
\label{sect:conclusions}

In this article we presented a systematic way to obtain an ordinary differential equation of minimal order for a given Feynman integral
in a given external variable.
The method reduces the problem of finding the differential equation to a problem of solving a linear system of equations.
The method works in $D$ dimensions and can be useful for the calculation of higher-order corrections in quantum field theory.
The inspiration for our method comes from algebraic geometry: Feynman integrals can be viewed as periods of mixed Hodge structures, which
vary with the external variables.
This variation is described by a Picard-Fuchs equation. With a suitable ansatz we can find the Picard-Fuchs equation, which in turn
is the sought-after differential equation for the Feynman integral.
When working in dimensional regularisation we observe interesting 
factorisation properties of the $D$-dimensional Picard-Fuchs operator when $D$ is specialised to integer dimensions.
This factorisation property is helpful when the differential equation is solved term by term as a Laurent series in the dimensional regularisation
parameter.

\bibliography{/home/stefanw/notes/biblio}

\begin{thebibliography}{10}

\bibitem{Kotikov:1990kg}
A.~V. Kotikov,
\newblock Phys. Lett. {\bf B254}, 158 (1991).

\bibitem{Kotikov:1991pm}
A.~V. Kotikov,
\newblock Phys. Lett. {\bf B267}, 123 (1991).

\bibitem{Remiddi:1997ny}
E.~Remiddi,
\newblock Nuovo Cim. {\bf A110}, 1435 (1997), hep-th/9711188.

\bibitem{Gehrmann:1999as}
T.~Gehrmann and E.~Remiddi,
\newblock Nucl. Phys. {\bf B580}, 485 (2000), hep-ph/9912329.

\bibitem{Gehrmann:2000zt}
T.~Gehrmann and E.~Remiddi,
\newblock Nucl. Phys. {\bf B601}, 248 (2001), hep-ph/0008287.

\bibitem{Gehrmann:2001ck}
T.~Gehrmann and E.~Remiddi,
\newblock Nucl. Phys. {\bf B601}, 287 (2001), hep-ph/0101124.

\bibitem{Argeri:2007up}
M.~Argeri and P.~Mastrolia,
\newblock Int. J. Mod. Phys. {\bf A22}, 4375 (2007), arXiv:0707.4037.

\bibitem{Tkachov:1981wb}
F.~V. Tkachov,
\newblock Phys. Lett. {\bf B100}, 65 (1981).

\bibitem{Chetyrkin:1981qh}
K.~G. Chetyrkin and F.~V. Tkachov,
\newblock Nucl. Phys. {\bf B192}, 159 (1981).

\bibitem{MullerStach:2011ru}
S.~M{\"u}ller-Stach, S.~Weinzierl, and R.~Zayadeh,
\newblock Commun. Num. Theor. Phys. {\bf 6}, 203 (2012), arXiv:1112.4360.

\bibitem{Hwa}
R.~C. Hwa and V.~L. Teplitz,
\newblock {\em Homology and Feynman Integrals} (W. A. Benjamin, 1966).

\bibitem{Boyling:1968}
J.~Boyling,
\newblock Nuovo Cim. {\bf A53}, 351 (1968).

\bibitem{Golubeva:1970gr}
V.~Golubeva,
\newblock Teor.Mat.Fiz. {\bf 3}, 405 (1970).

\bibitem{Lefschetz}
S.~Lefschetz,
\newblock {\em Applications of Algebraic Topology} (Springer, 1975).

\bibitem{Bloch:2006}
S.~Bloch, H.~Esnault, and D.~Kreimer,
\newblock Commun. Math. Phys. {\bf 267}, 181 (2006), math.AG/0510011.

\bibitem{Eden}
R.~J. Eden, P.~V. Landshoff, D.~I. Olive, and J.~C. Polkinghorne,
\newblock {\em The Analytic S-Matrix} (Cambridge University Press, 1966).

\bibitem{Nakanishi}
N.~Nakanishi,
\newblock {\em Graph Theory and Feynman Integrals} (Gordon and Breach, 1971).

\bibitem{Todorov}
I.~T. Todorov,
\newblock {\em Analytic Properties of Feynman Diagrams in Quantum Field Theory}
  (Pergamon Press, 1971).

\bibitem{Zavialov}
O.~I. Zavialov,
\newblock {\em Renormalized Quantum Field Theory} (Kluwer, 1990).

\bibitem{Itzykson:1980rh}
C.~Itzykson and J.~B. Zuber,
\newblock {\em Quantum Field Theory} (McGraw-Hill, New York, 1980).

\bibitem{Smirnov:2006ry}
V.~A. Smirnov,
\newblock {\em Feynman integral calculus} (Springer, Berlin, 2006).

\bibitem{Bogner:2010kv}
C.~Bogner and S.~Weinzierl,
\newblock Int. J. Mod. Phys. {\bf A25}, 2585 (2010), arXiv:1002.3458.

\bibitem{Griffiths:1969}
P.~A. Griffiths,
\newblock Ann. of Math. {\bf 90}, 460 (1969).

\end{thebibliography}
\bibliographystyle{/home/stefanw/latex-style/h-physrev5}

\end{document}